\documentclass[aps,prl,showpacs,floatfix,superscriptaddress,twocolumn]{revtex4}
\usepackage{graphicx}
\usepackage{amsmath}
\usepackage{amsfonts}
\usepackage{epstopdf}

\newcommand{\fref}[1]{(\ref{#1})}

\begin{document}

\title{Traffic jams, gliders, and bands in the quest for collective motion}

\author{Fernando Peruani} 
\affiliation{Max Planck Institute for the Physics of Complex Systems, N\"othnitzer Str. 38, 01187 Dresden, Germany}
\author{Tobias Klauss}
\affiliation{Zentrum f\"ur Informationsdienste und Hochleistungsrechnen, Technische Universit\"at Dresden, Zellescher Weg 12, 01069 Dresden, Germany}
\author{Andreas Deutsch}
\affiliation{Zentrum f\"ur Informationsdienste und Hochleistungsrechnen, Technische Universit\"at Dresden, Zellescher Weg 12, 01069 Dresden, Germany}
\author{Anja Voss-Boehme}
\affiliation{Zentrum f\"ur Informationsdienste und Hochleistungsrechnen, Technische Universit\"at Dresden, Zellescher Weg 12, 01069 Dresden, Germany}

\date{\today}

\begin{abstract}
We study a simple swarming model on a two-dimensional lattice where the self-propelled particles exhibit a tendency 
to align ferromagnetically. 
Volume exclusion effects are present: particles can only hop to a neighboring node if the node is 
empty.
Here we show that such effects lead to a surprisingly rich variety of
self-organized spatial
patterns. 
As particles exhibit an increasingly higher tendency to align to neighbors,
they first self-segregate into disordered particle aggregates. Aggregates turn
into {\it traffic jams}. 
Traffic jams evolve toward {\it gliders}, triangular high density regions that
migrate in a well-defined direction. 
Maximum order is achieved by the formation of elongated high density regions -
{\it bands} - that transverse the entire system. 
%
Numerical evidence suggests that below the percolation density the phase
transition associated to orientational order is of first-order, while at full
occupancy it is of second-order.
\end{abstract}

\maketitle


Self-propelled particle (SPP) systems are found at all scales in nature. 
%
Examples in biology range from human crowds~\cite{humans} and animal groups~\cite{cavagna10,bhattacharya10}, down to
insects~\cite{buhl06,romanczuk09}, bacteria~\cite{zhang10}, and 
even to the microcellular scale  with, e.g., the collective motion of microtubules 
driven by molecular motors~\cite{schaller10}.
SPP systems are not restricted to living systems. 
There are examples in non-living matter, as for instance in driven
granular media~\cite{narayan07,kudrolli08,deseigne10}. 
%
%
Interestingly, the statistical properties of the large-scale self-organized
patterns emerging in SPP systems depend only on a few microscopic details:  
the symmetry associated to the self-propulsion mechanism of the particles, which can be either {\it 
polar}~\cite{vicsek1995,gregoire2004,peruani2008} or {\it apolar}~\cite{chate2006},  
the symmetry of the velocity alignment mechanism, which can be either {\it ferromagnetic}~\cite{vicsek1995,gregoire2004}   
or {\it nematic}~\cite{chate2006,peruani2008}, 
and very importantly, the presence or absence of volume exclusion effects, as we are going to discuss here. 
In addition, the nature of the supporting space where the particles move plays also a crucial role: the 
dimension of the space, and whether this space is continuous~\cite{vicsek1995,gregoire2004,chate2006,peruani2008} or discrete~\cite{bussemaker1997,csahok1995,loan1999,raymond2006}.

\begin{figure*}
\centering\resizebox{17cm}{!}{\rotatebox{0}{\includegraphics{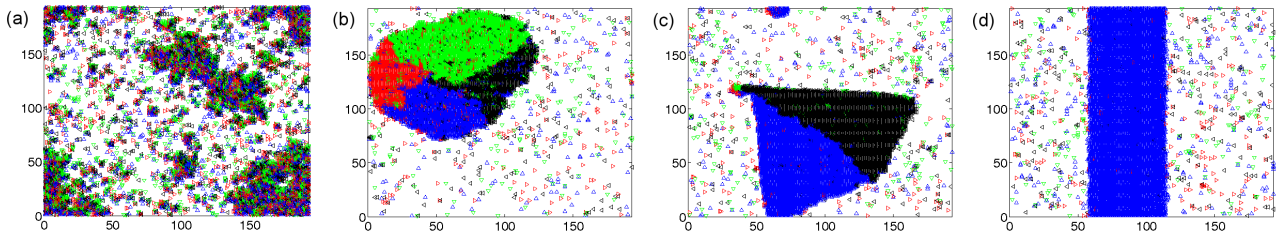}}}
\caption{(Color online) Example of self-organized spatial patterns: (a) disordered
  aggregates, (b) traffic jams, (c) gliders, and (d) bands. 
Particle orientation is
  indicated by the orientation of the small triangles, and is also
  color-coded: right (red), left (black), down (green), and up
  (blue). Parameter values are indicated in Fig.~\ref{fig:m_vs_g}. 
%
%
%
%
} \label{fig:snapshot_m_vs_g}
\end{figure*}

In this letter, we focus on {\it polar} SPPs moving on a two-dimensional lattice that align their 
orientation, respectively their moving direction, via a local {\it ferromagnetic} alignment 
mechanism.   
We explicitly model volume exclusion effects: nodes can be occupied at most by one particle. 
%
%
We show that  such effects 
introduce a coupling between particle speed, local density and local alignment
that lead to a surprisingly rich variety of
self-organized spatial patterns unseen in previous swarming models. 
As particles exhibit an increasingly higher tendency to align to neighbors, 
the system passes through three distinct phases. 
For weak alignment strength, the system exhibits orientational disorder,
while particles self-segregate, Fig.~\ref{fig:snapshot_m_vs_g}(a). 
Within this initial phase, there is a transition from a spatially homogeneous to
an aggregate phase. 
The onset of orientational (polar) order marks the beginning of the second
phase which is characterized by the emergence of 
locally ordered, high density regions: {\it traffic jams}, Fig.~\ref{fig:snapshot_m_vs_g}(b). 
As the tendency to align is enhanced, traffic jams evolve toward triangular
high density aggregates that migrate in a well-defined direction. 
We refer to these dynamical traffic jams as {\it gliders}, Fig. \ref{fig:snapshot_m_vs_g}(c). 
The third phase emerges when the particles self-organize into highly ordered, elongated, high
density regions: {\it bands}, Fig.~\ref{fig:snapshot_m_vs_g}(d). 
In contrast to the traveling bands observed in off-lattice SPP models with ferromagnetic
alignment~\cite{gregoire2004}, these bands are formed by particles aligned to the long axis of the
band and are rather static. 

We find evidence that the phase transition to orientational order is discontinuous below the percolation threshold. 
When the lattice is fully occupied, the system reduces to the
classical planar Potts model and the phase transition to orientational order is
undoubtedly of second-order~\cite{wu1982, betts1964}. 
Previous lattice swarming models were found to exhibit a continuous phase transition from a 
homogeneous to a condensed phase in 1D~\cite{loan1999,raymond2006}, while in 2D, both, first and 
second-order transitions to orientational order have been claimed. 
Bussemaker et al. reported a second-order transition in a cellular automaton model with 4 moving directions~\cite{bussemaker1997}, while Csah\'ok and Vicsek found, for a lattice-gas model with 6 moving directions, a weakly-first-order transition to collective migration~\cite{csahok1995}.   
Here we show that all these phenomena occur in a minimal 2D lattice
swarming model, but where in contrast to previous models the system dynamics is dominated
by volume exclusion effects that lead to a completely novel spatial
self-organization of particles.



\vspace{0.1cm} \noindent {\it Model.}
Particles move on a two-dimensional lattice with periodic boundary conditions, and have four possible orientations: up, down, 
left, and right, whose associated vectors are $\mathbf{v}_1$,  $\mathbf{v}_2$, $\mathbf{v}_3$, and 
$\mathbf{v}_4$, respectively.
The state of a particle is given by its position on the lattice and its orientation. As it will 
become clear below, the orientation of a particle fully determines its moving direction.  
%
%
The particles are able to perform two actions: i) they can change their orientation, and ii) they can migrate in the 
direction given by their orientation.
%
These actions have associated transition rates which specify the average number of events per time unit.
%
Let us start out with the reorientation transition rate $T_R$ that a particle at  
$\mathbf{x}$ with orientation $\mathbf{v}$ turns its orientation into direction $\mathbf{w}$:
\begin{equation}\label{eq:trans_turn}
T_R \left((\mathbf{x},\mathbf{v}) \to  (\mathbf{x},\mathbf{w})\right) = \exp(g 
\sum_{\mathbf{y} \, \in A(\mathbf{x})} \langle \mathbf{w} | \mathbf{V}(\mathbf{y})  \rangle ) \, ,
\end{equation}
where the sum runs over the nearest lattice neighbors of $\mathbf{x}$, represented by $A(\mathbf{x})$. The 
vector $\mathbf{V}(\mathbf{y})$ returns the orientation  of the particle placed in node 
$\mathbf{y}$, if there is any, and the null vector otherwise. The symbol $\langle . | . 
\rangle$ indicates the inner product between two vectors, while  $g$ is a parameter which 
controls the alignment {\it sensitivity}. 
For positive $g$, eq.~\fref{eq:trans_turn} defines a stochastic {\it ferromagnetic} alignment 
mechanism. 
As result of this alignment, first nearest neighbor particles tend to be aligned. 
%

The migration rate is defined in the following way:
\begin{equation}\label{eq:trans_mig}\begin{split}
T_M &\left((\mathbf{x},\mathbf{v}) \to  (\mathbf{y}=\mathbf{x}+\mathbf{v},\mathbf{v})\right) \\ 
 & =
 \left\{
\begin{array}{lcl}
v_0 & \mbox{if node $\mathbf{y}$} & \mbox{is empty}  \\
0 & \mbox{if node $\mathbf{y}$} & \mbox{is occupied.} 
\end{array} \right. 
\end{split}\end{equation}
Thus, a particle at position $\mathbf{x}$ and pointing in direction $\mathbf{v}$ migrates to the 
neighboring node  $\mathbf{y}=\mathbf{x}+\mathbf{v}$ with a transition rate $v_0$ as long as the node 
at $\mathbf{y}$ is empty. If the $\mathbf{y}$-node is occupied, the particle will not jump. 
The only action that is allowed to the particle in this situation is to change its orientation. 

At this point, it is important to understand how this continuous-time 
process is simulated. 
Let us assume that at time $t_0$ the system is in a given state. 
Then, we compute the time at which the next event will take place in the system, i.e., we calculate 
$t_1$. 
Now we have to decide which is the event that will take place at time $t_1$. 
We choose at random one out of all the possible events that could take place, but we weight each of 
these events  according to their associated transition rate, eqs.~\fref{eq:trans_turn} 
and~\fref{eq:trans_mig}. 
This procedure is an adaption of the classical Gillespie algorithm~\cite{gillespie77} to interacting particle 
systems~\cite{klauss2008}.

%
\begin{figure}
\centering\resizebox{5.5cm}{!}{\rotatebox{0}{\includegraphics{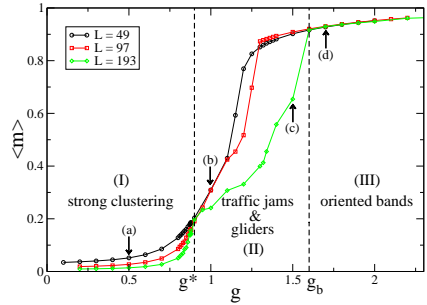}}}
\caption{
(Color online) 
Mean orientation $\langle m \rangle$ vs. the alignment sensitivity $g$ for systems with
density $d=0.3$ and migration rate $v_0=100$. 
Simulations were carried out for
$2\,10^7$ time steps. 
The different curves correspond to different system sizes $L$. 
%
The boundary between phase I and II, $g^*$, and the between phase II and III,
$g_b$ are indicated by the vertical dashed lines. 
(a) to (d) refer to the simulation snapshots shown in Fig.~\ref{fig:snapshot_m_vs_g}.
} \label{fig:m_vs_g}
\end{figure}

\begin{figure}
\centering\resizebox{8.0 cm}{!}{\rotatebox{0}{\includegraphics{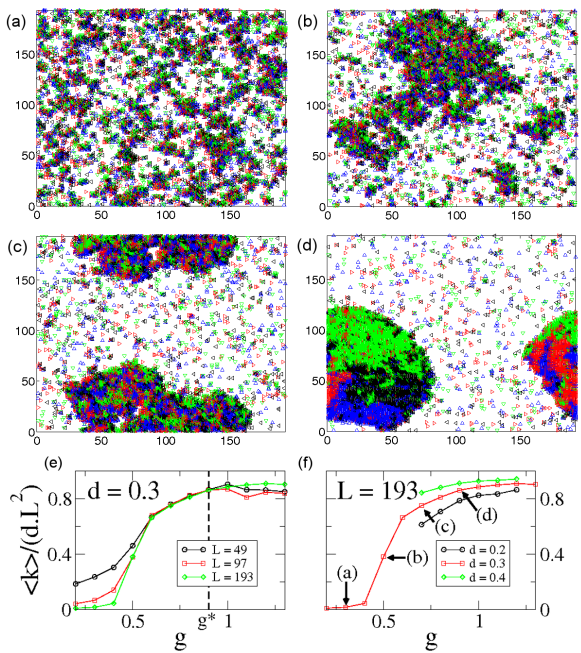}}}
\caption{(Color online) Phase transition to aggregation (phase I). (a) to (d) correspond to simulation snapshots whose value of $g$ is indicated in (f) (other parameters as in Fig.~\ref{fig:m_vs_g}). 
The average cluster size $\langle k \rangle$ normalized by the total number of particles, $d.L^2$, as function of $g$ is shown in (e) for various system sizes, and in (f) for various densities. } \label{fig:phaseA}
\end{figure}

\begin{figure}
\centering\resizebox{7cm}{!}{\rotatebox{0}{\includegraphics{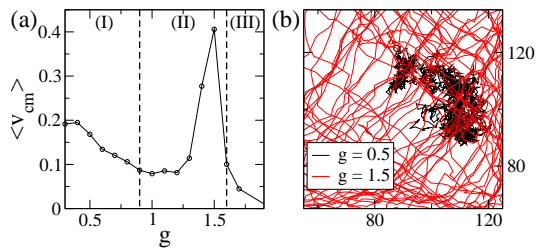}}}
\caption{(Color online) (a) average speed of the center of mass $\langle V_{cm} \rangle$  as function of $g$. (b) Trajectories of the center of mass for two different values of $g$. } \label{fig:trajectories}
\end{figure}

\vspace{0.1cm} \noindent {\it Results.} 
We start by fixing the migration rate $v_0$ and particle density $d$, and use
$g$ as control parameter. 
The degree of orientational order in the system at time $t$ is characterized by the (global) orientation  
 $m(t) = (1/N)|\sum_{\mathbf{x}} \mathbf{V}(\mathbf{x})|$, where $N$ is the total number of 
particles in the system, the sum runs over all lattice sites, and $\mathbf{V}(\mathbf{x})$ is defined as above.
Fig. \ref{fig:m_vs_g} shows the behavior of the mean orientation $\langle m \rangle$ as
function of the alignment sensitivity $g$, with $\langle \dots \rangle$ a
temporal average taken once the system reaches the steady-state after a short transient. 
The system exhibits a phase transition to orientational order above a critical $g^*$ for all
densities, as long as $v_0 > 0$.  
Here we focus on $0<d<d_p$, where $d_p$ refers to the (site) percolation threshold
in a 2D (square) lattice, $d_p \sim 0.59$. 
The system exhibits three phases with $g$, labeled
I, II, and III by increasing alignment sensitivity $g$.
%
%
%
Phase I corresponds to $g<g^*$ and is characterized by exhibiting  no
macroscopic orientational order. 
Fig. \ref{fig:phaseA} shows that within phase I there is a dynamic phase transition from
a spatially homogeneous to an aggregate phase as $g$ is
increased. 
The degree of aggregation is characterized by the average cluster size $\langle
k \rangle$, which is computed as the temporal average of $\langle
k \rangle =\sum_k k\,p(k,t)$, with $p(k,t) = k\,n_k(t)/N$, where $n_k(t)$ is the number of
clusters of mass $k$ at time $t$ and $N=d\,L^2$ is the number of particles in the
system. 
The figure shows that there exists a critical value $g_a$ above which a phase  transition to aggregation
occurs, with $g_a<g^*$. 
As $g$ approaches $g^*$,  more than $85$\% of the particles in
the system form a large aggregate.
%
%
%
%
%
\begin{figure}
\centering\resizebox{\columnwidth}{!}{\rotatebox{0}{\includegraphics{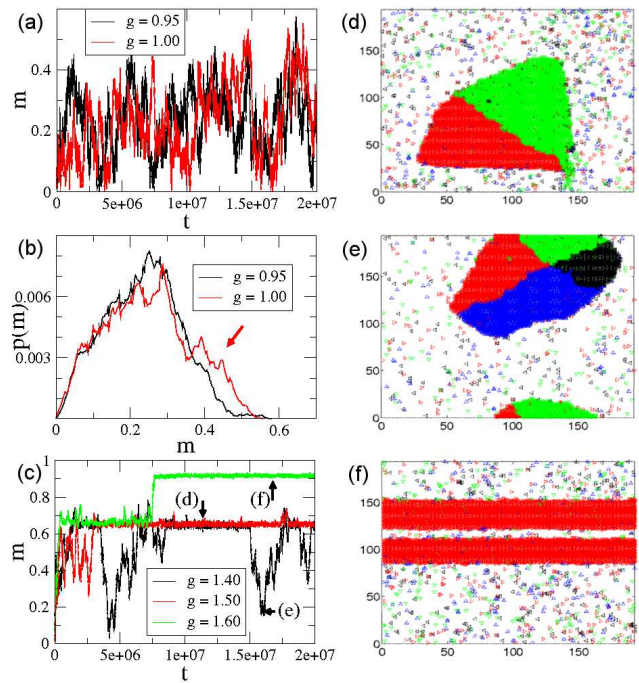}}}
\caption{
(Color online) Phase transition to orientational order. Time series $m(t)$, histogram, and typical spatial configurations for phase II, $g^* < g \leq g_b$. Letters in (c) indicate simulation snapshots shown in (d)-(f). Parameters as in Fig.~\ref{fig:m_vs_g}. 
%
%
%
%
} \label{fig:phaseB}
\end{figure}
The transition point between phase I and II is at $g^*$, where 
the curves $\langle m \rangle(g,L)$  for different system sizes $L$ meet.
Interestingly, $g^*$  seems to be independent of the density $d$, as
confirmed with simulations for various system sizes with density $d=0.2$, $0.3$,
and $0.4$ (data not shown). 
Moreover, $g^*$ coincides with the critical point for the full occupancy
problem, i.e., $d=1$. 
Fig. \ref{fig:phaseB}(a) shows time series of $m(t)$ for values of $g$ close
to $g^*$. The order parameter $m(t)$ exhibits fluctuations between 
high and low values. Low $m$-values correspond to the appearance of round traffic jams,
while high values correspond to elongated traffic jams where two directions dominate over the other two.
Traffic jams results from the jamming of  
four  particle clusters attempting to move to the left, right, upward, and
downward, respectively. 
Fluctuations are due to the competition between these four clusters. 
Fig.~\ref{fig:phaseB}(b) shows that the distributions $p(m)$ obtained from the $m(t)$ time series for values of $g$ close to $g^*$ do not exhibit a
Gaussian shape as expected for a second order transition, but rather a bimodal
distribution (the arrow indicates the second peak) as expected for first order transitions. 
The coexistence of several particle configurations (or ``phases'') exhibiting 
different degrees of ordering, i.e., values of $m$, is evident for values of
$g$ deep into phase II. 
Fig. \ref{fig:phaseB}(c) shows that $m(t)$ jumps between well-defined values corresponding to different spatial patterns.  
Values of $m(t)\sim 1/\sqrt{2}$ are associated to gliders, while higher values of $m(t)$ correspond to bands.  Lower values of $m(t)$ are due to the presence of traffic jams. 
Gliders are dynamical traffic jams moving backwards with respect to their mean average orientation. 
Their presence affects the temporal evolution of the center of mass of the system, $\mathbf{x}_{cm}(t)$, which exhibits ballistic motion whenever there is a glider, while otherwise is Brownian. As result of this, the average speed of the center of mass $\langle V_{cm} \rangle$ peaks at values of $g$ where gliders are more stable~\cite{def_speed_cm}, see Fig.~\ref{fig:trajectories}. 
Gliders are remarkably different from traffic jams 
observed in 2D traffic models~\cite{trafficjams2D}, arguably due to the
presence of the alignment mechanism, Eq.~(\ref{eq:trans_turn}). 
%
%
How frequently gliders appear and for how long they survive, depends on the value of $g$ and $L$. 
For example, for $g=1.4$,  $m(t)$ displays excursions from low to high
values that reflect the fact the system alternates between traffic jams and gliders.  
For an illustration of this dynamics, see~\cite{movie}.
For $g>g_b$, i.e. in phase III, the only stable configurations is a band. 
In contrast to $g^*$, $g_b$ is highly dependent on $L$. 
%
%

%

%
%

\vspace{0.1cm} \noindent {\it Discussion.}
In the limit of full occupancy, $d=1$,   
particles are frozen in their positions and the only action allowed to them is reorientation. 
The system defined by Eqs. \fref{eq:trans_turn} and \fref{eq:trans_mig} becomes an equilibrium 
system in this limit, whose order is again characterized by $m(t)$.  
Since, by definition, there are only four possible orientations, we can safely claim that the model 
reduces to a 4-state planar Potts model~\cite{wu1982}.
It has been shown that this model can be reduced to the standard $2$-Potts 
model~\cite{betts1964}. 
In two-dimensions, the standard $q$-Potts model exhibits a continuous transition for $q \leq 4$, and 
a discontinuous one for $q>4$~\cite{wu1982}. 
Thus, our model exhibits a second-order transition for $d=1$, as confirmed via
simulations (data not shown). 
%
%
%
%
%
%
%
For $v_0>0$ and $d<1$, we are in a pure non-equilibrium scenario. 
The migration rule, Eq. \fref{eq:trans_mig}, breaks detailed balance and prevents us from writing down a free energy.  
Nevertheless, it is worth to compare our system with its equilibrium counterpart, the diluted Potts model with annealed vacancies. 
If we represent the absence of particles in a lattice position with an extra vector direction, we end up with a $5$-Potts system with full occupancy, instead of 
a diluted $4$-Potts system. In consequence, according to what was said above, the 
transition would be first-order. 
The argument applies to the standard Potts model and assumes vacancies are in
thermal equilibrium, which is not true for Eq. \fref{eq:trans_mig}. 
Nevertheless, it helps to realize that a discontinuous dynamic phase transition is quite 
possible in our non-equilibrium system.

In summary, we have shown through a minimal model that volume exclusion effects, when they are allowed to stop particle motion, can lead to a surprisingly rich variety of self-organized patterns. 
Such effects introduce a  coupling between local density, local orientation
and {\it particle speed} that strongly affects the large-scale behavior of the
system, with the jamming of particles playing a dominant role. 
This coupling is present in many  real systems as in gliding
bacteria, animal groups, etc.  
Certainly, several features of the self-organized patterns described here 
depend on the discrete nature of the model.  
Nevertheless, we expect similar phenomena to emerge in off-lattice, continuum symmetry systems.  
For instance, static traffic jams are probably a robust property of all systems where stagnation can
occur. 
Here we have also learned that the jamming of self-propelled particles can lead to unexpected self-organized structures in two-dimensions like dynamical traffic jams, e.g., gliders. 
%
The presence of an alignment mechanism induces (local) orientational order, and
provided particles are oriented, density waves of stagnated particles should
emerge. 
The results reported here are a first step toward a deep understanding of 
the possible phenomena that such a coupling may induce.

We thank A.~Greven and C.F.~Lee for useful comments.


\begin{thebibliography}{99}
\bibitem{def_speed_cm} $\langle V_{cm} \rangle$ has been computed as $\langle
  V_{cm} \rangle = \langle | \mathbf{x}_{cm}(t+\tau)-\mathbf{x}_{cm}(t)|/\tau
  \rangle_{t}$, with $\tau = 10^4$ and $\langle \hdots \rangle$ a temporal
  average.
\end{thebibliography}
\end{document}